%% file: paper.tex
\documentclass{iau}
\usepackage{graphicx}

\def\Mpch{~h^{-1} {\rm Mpc}}

\def\MpchVolume{~(h^{-1} {\rm Mpc})^3}

\def\Msun{\rm{M}_{\odot}}
\def\Msolar{~h^{-1} \rm{M}_{\odot}}

\newcommand{\Nexus}{\textsc{NEXUS}}
\newcommand{\nexus}{\textsc{NEXUS+}}

\newcommand{\reffig}[1]{Fig.~\ref{#1}}

\usepackage{natbib}

\title[Understanding the cosmic web]
{Understanding the cosmic web}

\author[Cautun et al.]
{Marius~Cautun$^{1,2}$,
Rien~van~de~Weygaert$^{2}$,
Bernard~J.~T.~Jones$^{2}$,
\and Carlos~S.~Frenk$^{1}$}

\affiliation{$^1$Department of Physics, Institute for Computational Cosmology, University of Durham, \\ South Road, Durham DH1 3LE, UK,  email: {\tt m.c.cautun@durham.ac.uk} \\
  $^2$Kapteyn Instituut, Rijksuniversiteit Groningen, P.O. Box 800,
  9700 AV Groningen, \\The Netherlands}

\pubyear{2014}
\volume{308} 
\jname{The Zeldovich Universe: Genesis and Growth of the Cosmic Web}
\editors{R.~van~de~Weygaert, S.~Shandarin, E.~Saar \& J.~Einasto, eds.}

\begin{document}

\maketitle

\begin{abstract}
We investigate the characteristics and the time evolution of the cosmic web from redshift, $z=2$, to present time, within the framework of the \nexus{} algorithm. This necessitates the introduction of new analysis tools optimally suited to describe the very intricate and hierarchical pattern that is the cosmic web. In particular, we characterize filaments (walls) in terms of their linear (surface) mass density. This is very good in capturing the evolution of these structures. At early times the cosmos is dominated by tenuous filaments and sheets, which, during subsequent evolution, merge together, such that the present day web is dominated by fewer, but much more massive, structures. We also show that voids are more naturally described in terms of their boundaries and not their centres. We illustrate this for void density profiles, which, when expressed as a function of the distance from void boundary, show a universal profile in good qualitative agreement with the theoretical shell-crossing framework of expanding underdense regions.
\end{abstract}

\section{Introduction}

\begin{figure}[tbp]
  \begin{center} 
    \includegraphics[width=\textwidth]{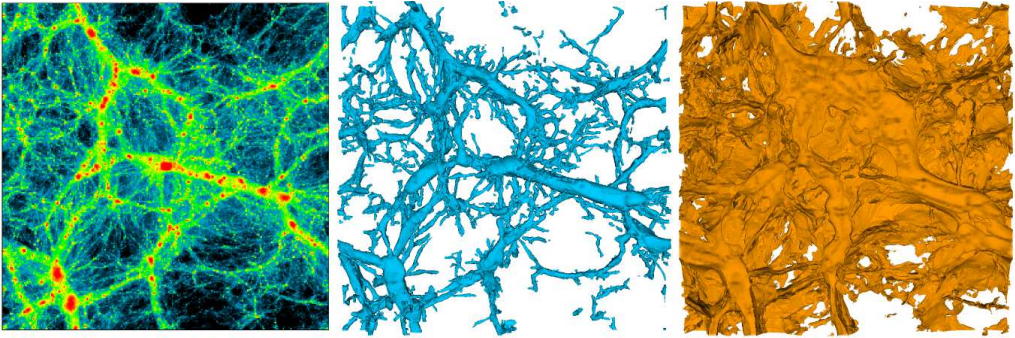}
  \end{center} 
  \caption{ The complexity and multiscale character of the cosmic web as identified by \nexus{}. It shows the density field (left), the filaments (centre) and the walls (right) in a $100\times100\times10\MpchVolume$ slice trough the Millennium-II Simulation. }
  \label{fig:cosmic_web}
\end{figure}

On megaparsec scales the matter distribution of the Universe is not uniform, but it forms an intricate pattern which is known as the \textit{Cosmic Web} \citep{Bond1996}. This is the most salient feature of the anisotropic gravitational collapse of matter, the motor behind the formation of structure in the Universe. Identifying and characterising the cosmic web network, in both numerical simulations and observations, is very challenging due to the overwhelming complexity of the individual structures, their connectivity and their intrinsic multiscale nature. It is even more difficult to follow the time evolution of the cosmic web, since the dominant components and scales change rapidly with redshift. This necessitates the use of scale- and user-free methods that naturally adapt to the complex geometry of the web and that extract the maximum information available among the components of this network.

\section{Simulation and methods} 

We follow the evolution of the cosmic web using the high resolution Millennium and Millennium-II dark matter simulations \citep{millSim,millSim2}, which describe the formation of structure in a periodic box of length $500$ and $100\Mpch$, respectively. We use the Delaunay Tessellation Field Estimator \citep{Schaap2000,2009LNP...665..291V,Cautun2011} to obtain continuous density and velocity fields. These are used as input for the \nexus{} algorithm \citep{nexus2013}, which identifies the cosmic web components. 
\nexus{} uses the mass distribution, as traced by the density field, to identify clusters, filaments, walls and voids. The dominant morphological signal is extracted from a 4-dimensional scale-space representation of the matter distribution. The fourth dimension is constructed by filtering the logarithm of the density with a Gaussian kernel, for a range of smoothing scales. The environments are classified on the basis of the dominant local morphological signature of the filtered density field. 

The outcome of applying \nexus{} to the density field is illustrated in \reffig{fig:cosmic_web}. 
Based on the Multiscale Morphology Filter (MMF) method of \citet{Aragon07b}, we developed \nexus{}, and its sister method, \Nexus{}, to use a multitude of tracer fields for classifying the cosmic web environments. It makes use of the density, tidal, velocity shear and velocity divergence fields. In addition, \Nexus{} and \nexus{} have the unique feature of employing natural and self-consistent criteria for identifying the cosmic web, an improvement with respect to the less clear percolation threshold criteria used in MMF.
This leads to approaches optimally suited for identifying the morphological environments since these methods are multiscale, parameter-free and designed to fully account for the anisotropic nature of gravitational collapse.

For this study, \nexus{} has two main advantages. Firstly, it determines in a self-consistent way all the different morphological components. And, secondly, the multiscale character of the method makes it ideal not only for identifying prominent as well as tenuous environments, but also for studying the time evolution of the cosmic web without having to choose a user-defined scale.

\section{The evolution of the cosmic web}

\begin{figure}[tb]
  \centering
    \includegraphics[width=.85\textwidth]{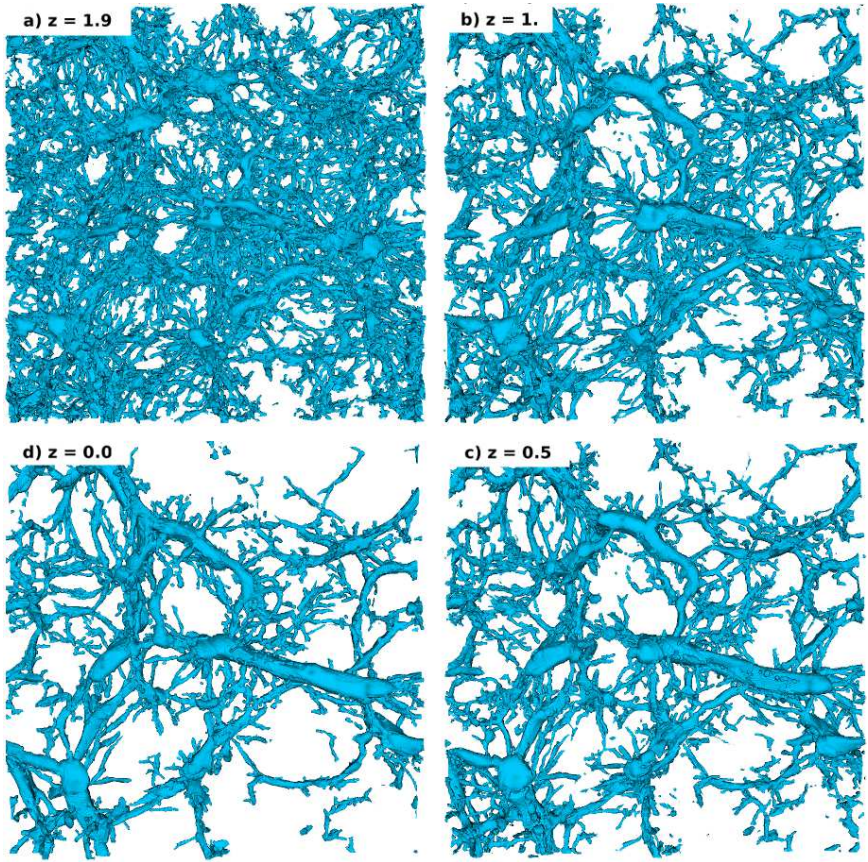}
  \caption{ The evolution of the filamentary network as identified by \nexus{} in a $100\times100\times10\MpchVolume$ slice trough the Millennium-II Simulation. }
  \label{fig:fila_evolution}
\end{figure}

\begin{figure}[tbp]
  \begin{center} 
    \includegraphics[width=.4\textwidth, angle=-90]{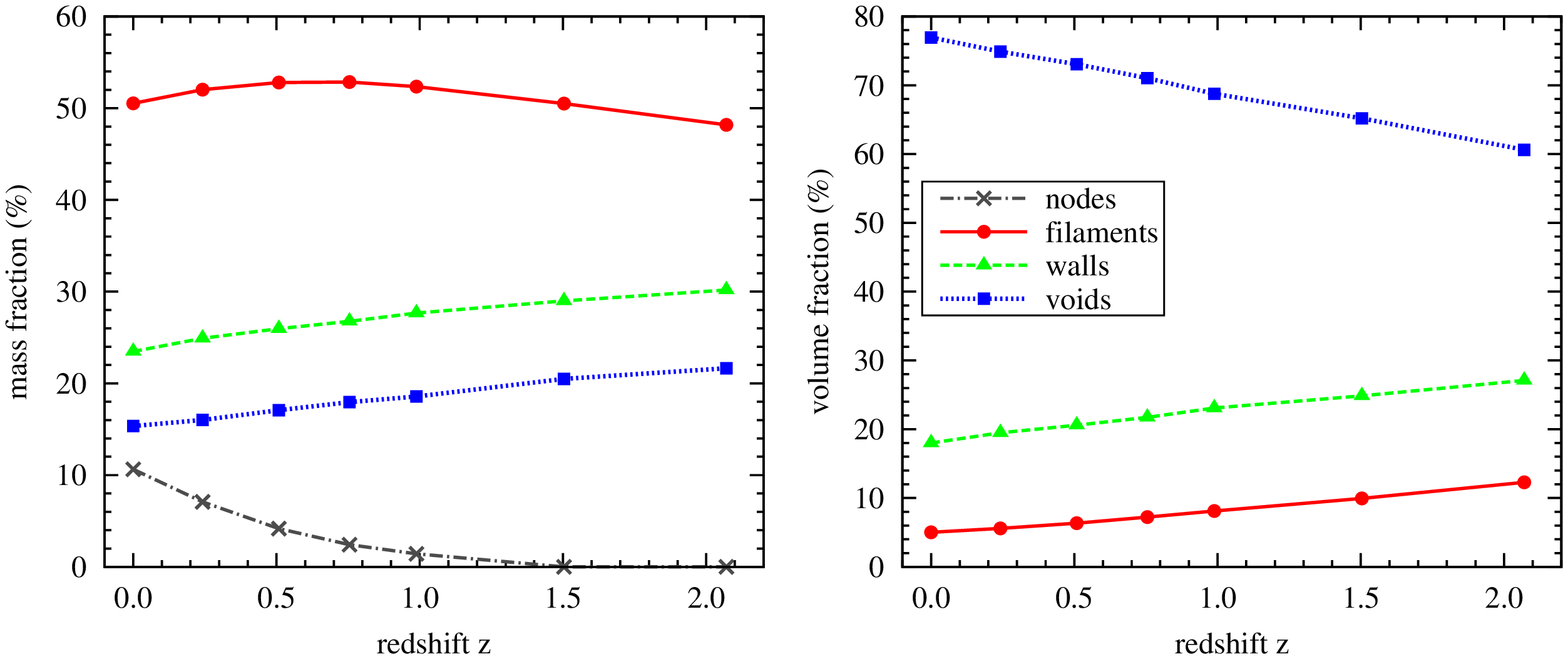} \\
    \includegraphics[width=.4\textwidth, angle=-90]{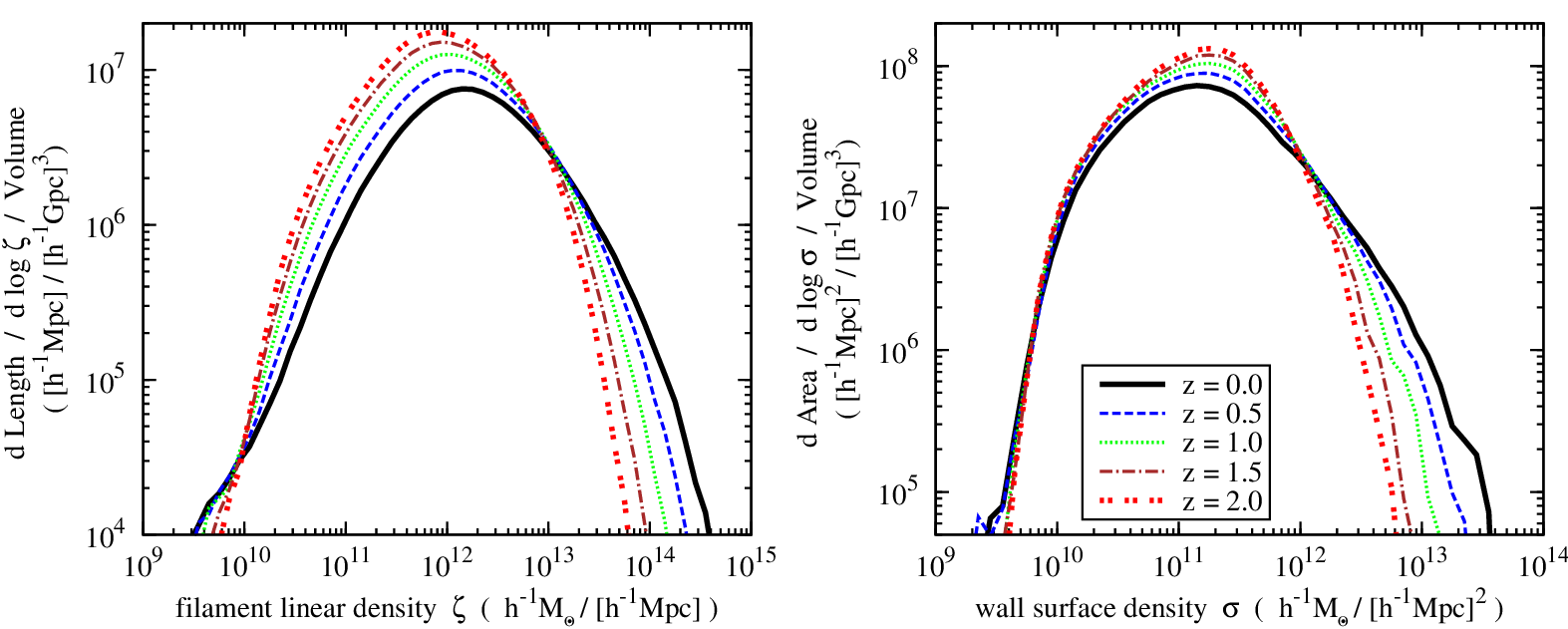}
  \end{center} 
  \caption{\textit{Top row:} the time evolution of the mass and volume fraction in each cosmic web environment. \textit{Bottom row:} the distribution of linear mass density of filaments (left) and the surface mass density of walls (right) at different redshifts. }
  \label{fig:fila_prop}
\end{figure}

\reffig{fig:fila_evolution} shows the evolution of filamentary environments starting with a redshift of $z=1.9$ down to the present time. At early times, the filaments form a complex network that pervades most of the cosmic volume, with the exception of the most underdense regions. While the network has a few thick structures, it is dominated by small scale filaments. These thin filaments seem to be packed much more tightly close to prominent structures, suggesting that overdense regions have a higher richness of filaments. By $z=1$, we find that most of the tenuous structures have disappeared and that we can more easily see the pronounced filaments. 
Going forward in time, to $z=0.5$ and $0$, we find that the evolution of the cosmic web significantly slows down, with only minor changes after $z=0.5$. 
Though not shown, a similar evolution can be seen for the wall network too. For a more in depth analysis see \citet{Cautun2014}.

\subsection{The mass distribution along filaments and walls}
One of the most widely employed methods to study the cosmic web properties involve the use of global properties, like mass and volume filling fraction of each component \citep[e.g.][]{Aragon07b,Hahn2007a,Forero-Romero09}. When applied to our study, such an approach leads to the results presented in top row of \reffig{fig:fila_prop}. For example, it shows that the volume occupied by filaments decreases since high redshift. This is probably due to the merging of the thin and tenuous filaments with the more prominent structures. However, such a simple analysis cannot characterise the complex evolution seen in \reffig{fig:fila_evolution}. For example, it cannot tell which components, tenuous or prominent, contained the most mass and how this mass distribution evolves in time. To do so, one needs a more complex analysis framework that takes into account the geometry of the various cosmic components.

Such a framework has been introduced in \citet[][see also \citealt{2010MNRAS.408.2163A}]{Cautun2014} and takes advantage that, to a first approximation, filaments and walls can be seen as 1-dimensional lines and 2-dimensional surfaces, respectively. These represent the spine of filaments and the central plane of sheets; and can be computed using the techniques described in \citet{2010MNRAS.408.2163A} and \citet{Cautun2014}. This reduces the complex filamentary network to a simpler distribution of interconnected curves, in which all the mass and haloes in the filament is compressed to its spine. Following this, one can move along the resulting curves and compute local quantities, like the mean linear density and the mean diameter of filaments \citep[for details see][]{Cautun2014}. 

The bottom-left panel of \reffig{fig:fila_prop} shows the results of such an analysis. It gives the length of filaments, per unit volume, that have a given linear mass density, when measured within a window of $2\Mpch$. The figure shows that there are few very low or very high mass filaments and that most of the length of the filamentary network is given by segments with linear densities of ${\sim}10^{12}\Msun/{\rm Mpc}$. This distribution evolves in time, to show that segments with high mass become more common at late times, while at the same time there are fewer low mass segments. Even more telling, is the shift to the right in the peak of the distribution, showing that time evolution leads to an increase in the mass of the typical filament segment. 

In contrast to filaments, the bottom-right panel of \reffig{fig:fila_prop} shows that the typical sheet regions become less massive at present time. The decrease in wall surface density is seen as the shift in the peak of the distribution towards lower $\sigma_{\rm wall}$ values at later times. It shows that the decrease in the mass fraction of walls seen in the top-left panel takes place via two processes. First, as we just argued, typical sheet stretches become less massive. And secondly, the extent of the wall network reduces at later times, as seen in the decreasing peak values of the $\sigma_{\rm wall}$ distribution.

\subsection{The halo distribution among the web environments}
\begin{figure}[tb]
  \centering
    \includegraphics[width=.38\textwidth, angle=-90]{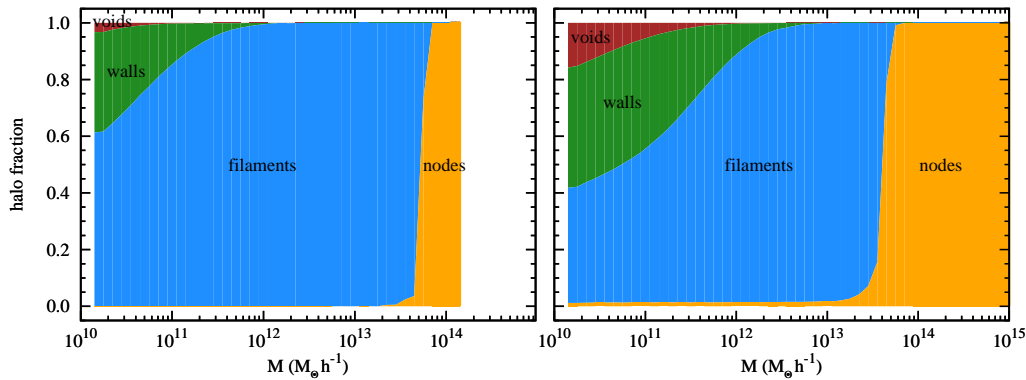}
  \caption{ The fraction of haloes in each cosmic web environment as a function of halo mass. We shows results for redshift, $z=2$ (left), and for the present time (right). }
  \label{fig:evolution_halo_frac}
\end{figure}
\begin{figure}[tb]
  \centering
    \includegraphics[width=.95\textwidth]{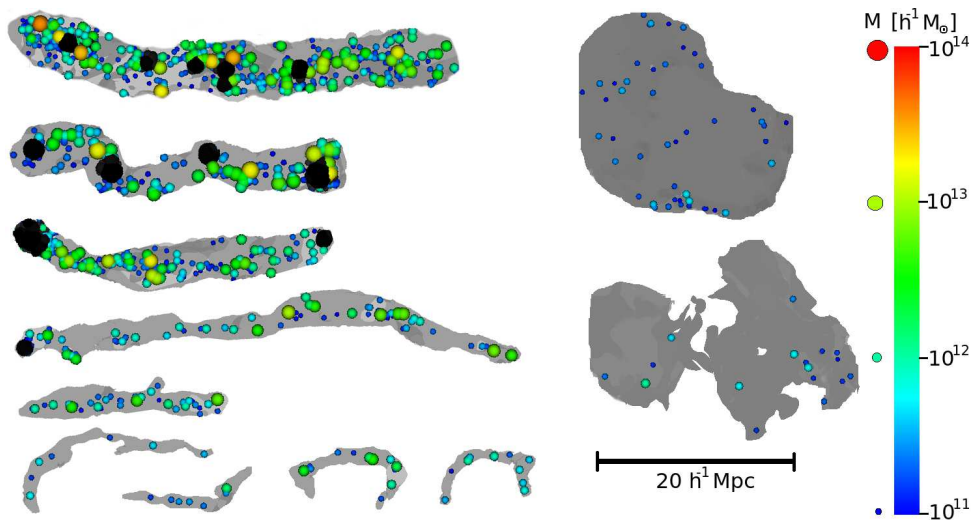}
  \caption{ The haloes populating a few typical cosmic web filaments (left) and walls (right). The colour and size of the haloes indicates their mass (see legend on the right) and the black points show haloes found in node environments.}
  \label{fig:evolution_halo_populations}
\end{figure}

The distribution of haloes across web environments plays a key role. First, the morphology of the cosmic web determines the preferential directions of accretion and thus influences the shape and angular momentum of haloes, inducing large scale alignments \citep[e.g.][]{Aragon07a,Hahn2007a}. And secondly, to identify and study the cosmic web in the observational reality one makes use of galaxies, which are hosted in haloes. So knowing how haloes populate the morphological components is crucial to understand the web environments seen in observations. To this end, we show in \reffig{fig:evolution_halo_frac} the distribution of haloes in the cosmic web at both $z=2$ and 0. The figure shows a clear segregation of haloes across environments, with the most massive ones living in nodes and prominent filaments. Walls typically host ${\sim}10^{12}\Msolar$ and lower mass objects, while void regions are populated with even lower mass haloes.

More interestingly, the halo population varies not only with environment, but also between structures with similar morphological features. This is shown in the left panel of \reffig{fig:evolution_halo_populations}, where we illustrate the haloes found in several filamentary branches. From top to bottom, we show progressively less prominent filamentary branches, with the bottom most ones corresponding to filaments found in voids. The panel shows a clear trend of the halo distribution with filament properties. Thicker filaments, which are typically outstretched between cluster pairs, are populated with more massive haloes which are also more tightly packed together. In contrast, haloes in tenuous filaments are typically low mass, similar to the ones in walls, and are widely spaced apart. 

\reffig{fig:evolution_halo_populations} explains why prominent filaments are easy to find observationally: they host many bright galaxies. In contrast, tenuous structures are mostly inhabited by a dilute population of low luminosity galaxies. Detecting these structures presents many observational challenges since such tenuous objects are hardly conspicuous in the spatial distribution of galaxies. The configuration of three aligned galaxies inside a void found by \citet[][]{Beygu2013} is probably an example of such a thin filament \citep{Rieder2013}.

\subsection{The mass transport across the cosmic web}
\begin{figure}[tb]
  \centering
    \includegraphics[width=\textwidth]{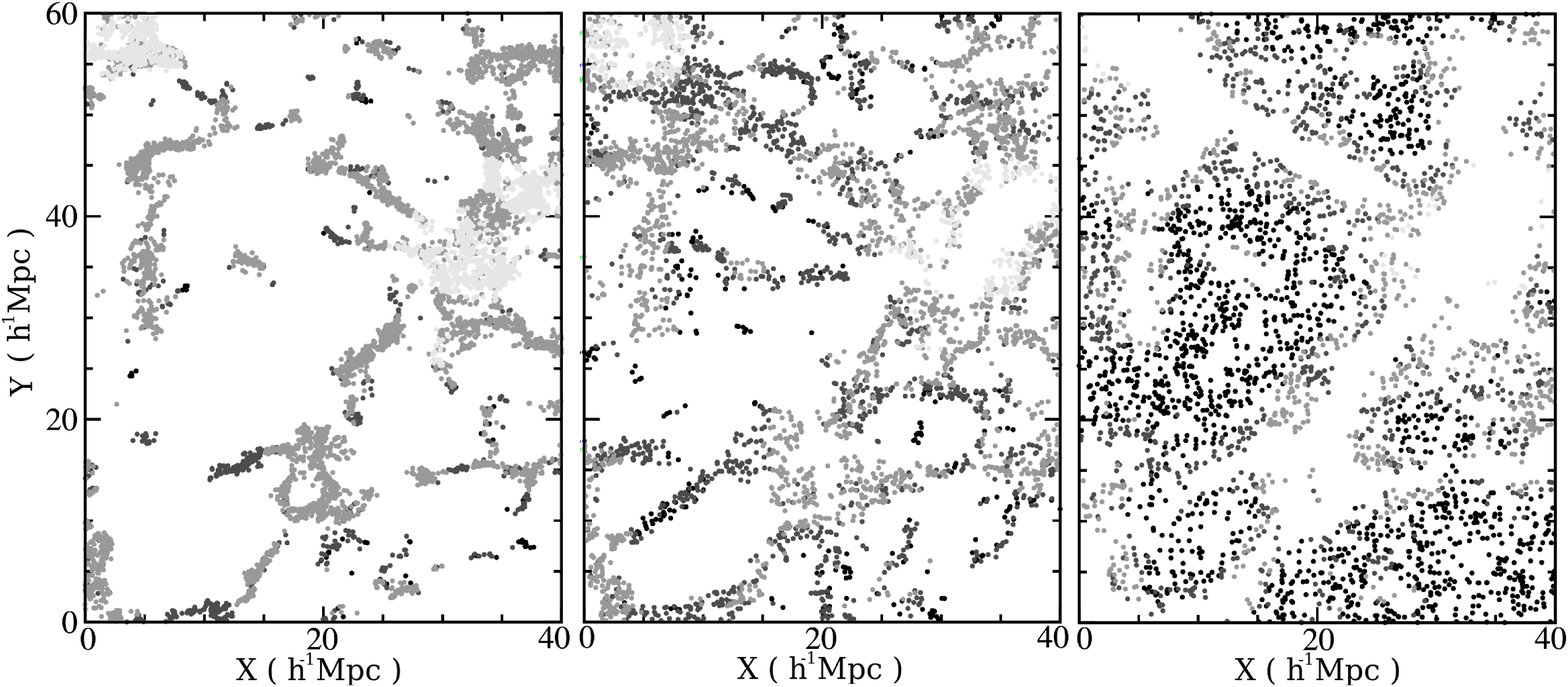}
  \caption{ The mass transport across the web components. It shows dark matter particles identified at $z=2$ as part of filaments (left), walls (centre) and voids (right). These particles are coloured according to their web classification at $z=0$: node (light-grey), filament (grey), wall (dark-grey) and void (black).}
  \label{fig:evolution_mass_transport}
\end{figure}

By comparing the cosmic web identified at different redshifts, we can trace the path taken by the anisotropic collapse of matter and study the transport of mass among the different web environments. According to the gravitational instability theory, the matter distribution follows a well defined path, with mass flowing from voids into sheets, from sheets into filaments and only in the last step into the cosmic nodes. We illustrate this with the help of \reffig{fig:evolution_mass_transport}, which shows the distribution of dark matter particles in a thin $2\Mpch$ slice at redshift $z=2$. Each panel gives the particles identified as part of filaments, walls and voids at $z=2$, with the particles coloured according to the environments they are found in at the present time.

\reffig{fig:evolution_mass_transport} shows some of the most important characteristics of the mass transport across the cosmic web. These conclusions are supported by the more in-depth and quantitative analysis of \citet{Cautun2014}. Among others, it shows that nodes form at the intersections of the filamentary network and that most of the mass in them has been accreted from regions that correspond to $z=2$ filaments. For filaments, most of the mass found in these objects at $z=2$ is also found in filaments at present time. In contrast, walls loose more than half of their $z=2$ mass to mostly filaments, with the rest remaining in walls. And finally, voids also loose around half of their $z=2$ mass, which flows into walls and filaments. In addition, \reffig{fig:evolution_mass_transport} illustrates some of the limitations of the analysis. A small fraction of $z=2$ filaments and walls are identified as present day walls and voids, respectively. This is restricted to minor filaments and walls and it is due to the difficulty of identifying such tenuous structures, especially across multiple time steps.

\section{A natural void profile}

\begin{figure}[tbp]
  \begin{center} 
    \includegraphics[width=.34\textwidth, angle=-90]{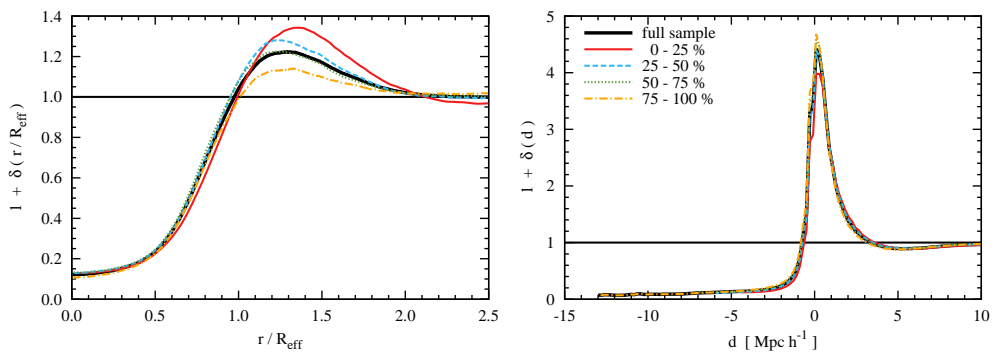}
  \end{center} 
  \caption{The density profile, $1+\delta$, of voids for the full population and for subsamples split according to the volume of the void, from small to large. The left panel show the spherical profile of voids. The right panel shows the void profile as a function of the distance from the void boundary. We use the convention that negative $d$ values correspond to the inside of voids, while positive ones correspond to the profile outside the void. }
  \label{fig:void_profile}
\end{figure}

The filaments and walls are not the only ones that have complex shapes and morphologies. Voids have it too. The simple picture of an expanding underdensity in a uniform background suggests that voids become more spherical as they evolve \citep{Icke1984,Shethwey04}. But in reality, voids are not isolated and this simple picture does not hold. There are two major factors that affect the evolution of voids. Firstly, since voids are nearly empty the force field that dictates their growth is dominated by external forces. This external tidal field determines the anisotropic expansion of voids and manifest itself as large scale correlations, over distances larger than $30\Mpch$, between the shapes of neighbouring voids \citep{Platen2008}. Secondly, as voids expand, they encounter neighbouring voids resulting in a packing problem. These effects lead to voids that have highly aspherical and complex shapes \citep{Platen07,Platen2008}. 

Up to now, voids were characterised in terms of spherical profiles with respect to the void's center, typically the barycentre. Such a methodology results in the void profiles shown in the left panel of \reffig{fig:void_profile}. These profiles have been rescaled following the prescription of \citet{Hamaus2014}, where $R_{\rm eff}$ is the effective radius of the void corresponding to a spherical void with the same volume as the real object. This results in void profiles that are not fully universal and where one needs to follow a complex procedure to rescale voids of different sizes \citep[for details see][]{Hamaus2014,Nadathur2014}. More worryingly, the resulting profile does not show the large density caustic at their boundary resulting from the shell-crossing of expanding underdense shells \citep{Shethwey04}. So what is the reason for the mismatch?

The discrepancy between spherical void profiles and the shell-crossing predictions of \citet{Shethwey04} arise from the fact that voids are very far from having a spherical shape. Thus, using spherically averaged profiles does not lead to an accurate description of void structure. In fact, it is easier to determine the void boundary, where most of the mass and galaxies reside, than the void center, which is devoid of tracers. In fact, this very fact is the cornerstone of Watershed-based void finders \citep{Platen07,Neyrinck2008}. Thus, it is more natural to describe voids with respect to their boundary than with respect to their center. 

It suggests that void profiles, and in general void properties, should also be computed with respect to the void boundary. For this, we define the boundary distance field $d_i(\mathbf{x})$ as the minimum distance between point $\mathbf{x}$ and the boundary of void $i$. In addition, to distinguish between points inside and outside the void, we take the convention that $d_i(\mathbf{x})$ is negative if $\mathbf{x}$ is inside void $i$ and positive otherwise. The resulting density profiles, as a function of the void boundary distance, are shown in the right panel of \reffig{fig:void_profile}. First, we find the sharp increase in the density profile at $d{\sim}0$, corresponding to the caustic resulting from shell-crossing. And secondly, voids of different size show a much more similar and universal profile. This method gives a better and more natural description of not only density profiles, but also of void velocity profiles (Cautun et al. 2015, in prep.). 

\section*{Acknowledgements}
MC and CSF acknowledge the support of the ERC Advanced Investigator grant COSMIWAY [grant number GA 267291].
RvdW acknowledges support by the John Templeton Foundation, grant [\#FP05136-O].

\input{references}

\end{document}

%% file: references.tex
\newcommand{\aj}{AJ}
\newcommand{\actaa}{Acta Astron.}
\newcommand{\araa}{ARA\&A}
\newcommand{\apj}{ApJ}
\newcommand{\apjl}{ApJ}
\newcommand{\apjs}{ApJS}
\newcommand{\ao}{Appl.~Opt.}
\newcommand{\apss}{Ap\&SS}
\newcommand{\aap}{A\&A}
\newcommand{\aapr}{A\&A~Rev.}
\newcommand{\aaps}{A\&AS}
\newcommand{\azh}{AZh}
\newcommand{\baas}{BAAS}
\newcommand{\caa}{Chinese Astron. Astrophys.}
\newcommand{\cjaa}{Chinese J. Astron. Astrophys.}
\newcommand{\icarus}{Icarus}
\newcommand{\jcap}{J. Cosmology Astropart. Phys.}
\newcommand{\jrasc}{JRASC}
\newcommand{\memras}{MmRAS}
\newcommand{\mnras}{MNRAS}
\newcommand{\na}{New A}
\newcommand{\nar}{New A Rev.}
\newcommand{\pra}{Phys.~Rev.~A}
\newcommand{\prb}{Phys.~Rev.~B}
\newcommand{\prc}{Phys.~Rev.~C}
\newcommand{\prd}{Phys.~Rev.~D}
\newcommand{\pre}{Phys.~Rev.~E}
\newcommand{\prl}{Phys.~Rev.~Lett.}
\newcommand{\pasa}{PASA}
\newcommand{\pasp}{PASP}
\newcommand{\pasj}{PASJ}
\newcommand{\qjras}{QJRAS}
\newcommand{\rmxaa}{Rev. Mexicana Astron. Astrofis.}
\newcommand{\skytel}{S\&T}
\newcommand{\solphys}{Sol.~Phys.}
\newcommand{\sovast}{Soviet~Ast.}
\newcommand{\ssr}{Space~Sci.~Rev.}
\newcommand{\zap}{ZAp}
\newcommand{\nat}{Nature}
\newcommand{\iaucirc}{IAU~Circ.}
\newcommand{\aplett}{Astrophys.~Lett.}
\newcommand{\apspr}{Astrophys.~Space~Phys.~Res.}
\newcommand{\bain}{Bull.~Astron.~Inst.~Netherlands}
\newcommand{\fcp}{Fund.~Cosmic~Phys.}
\newcommand{\gca}{Geochim.~Cosmochim.~Acta}
\newcommand{\grl}{Geophys.~Res.~Lett.}
\newcommand{\jcp}{J.~Chem.~Phys.}
\newcommand{\jgr}{J.~Geophys.~Res.}
\newcommand{\jqsrt}{J.~Quant.~Spec.~Radiat.~Transf.}
\newcommand{\memsai}{Mem.~Soc.~Astron.~Italiana}
\newcommand{\nphysa}{Nucl.~Phys.~A}
\newcommand{\physrep}{Phys.~Rep.}
\newcommand{\physscr}{Phys.~Scr}
\newcommand{\planss}{Planet.~Space~Sci.}
\newcommand{\procspie}{Proc.~SPIE}
\bibliographystyle{mn2e}
\bibliography{references}